# Two Perspectives on the Twist of DNA

by

Lauren A. Britton, Wilma K. Olson, and Irwin Tobias[a)]

*Department of Chemistry & Chemical Biology, BioMaPS Institute for Quantitative Biology, Rutgers, the State University of New Jersey, 610 Taylor Road, Piscataway, NJ 08854, USA*

## ABSTRACT

Because of the double-helical structure of DNA, in which two strands of complementary nucleotides intertwine around each other, a covalently closed DNA molecule with no interruptions in either strand can be viewed as two interlocked single-stranded rings. Two closed space curves have long been known by mathematicians to exhibit a property called the linking number, a topologically invariant integer, expressible as the sum of two other quantities, the twist of one of the curves about the other, and the writhing number, or writhe, a measure of the chiral distortion from planarity of one of the two closed curves. We here derive expressions for the twist of supercoiled DNA and the writhe of a closed molecule consistent with the modern view of DNA as a sequence of base-pair steps. Structural biologists commonly characterize the spatial disposition of each step in terms of six rigid-body parameters, one of which, coincidentally, is also called the twist. Of interest is the difference in the mathematical properties between this step-parameter twist and the twist of supercoiling associated with a given base-pair step. For example, it turns out that the latter twist, unlike the former, is sensitive to certain translational shearing distortions of the molecule that are chiral in nature. Thus, by comparing the values for the two twists for each step of a high-resolution structure of a protein-DNA complex, the

---

a) Electronic mail: tobias@rutchem.rutgers.edu

nucleosome considered here, for example, we may be able to determine how the binding of various proteins contributes to chiral structural changes of the DNA.

## INTRODUCTION

The structure of a DNA molecule is often described as a succession of base pairs, each represented as a rectangular plane.[1] A knowledge of the relative locations of origins positioned within these planes and the relative orientations of the short and long axes of the rectangles allows one to determine for each pair of adjacent base pairs in the molecule – a so-called base-pair step – the numerical values of six rigid-body parameters, three translational: shift, slide, and rise, and three angular: tilt, roll, and twist.[2-7]

Some forty years ago mathematicians defined a “twist” which, shortly after its introduction, was applied to DNA and used, like the step-parameter twist mentioned above, to characterize the secondary structure of the molecule.[8-10] This twist was defined as the value of a certain integral involving two continuous space curves. In the application to DNA, the structure of which at that time was often depicted in terms of space curves,[11] one of the curves was taken to be the axis of the double helix, and the other one of the strands winding about this axis. One then went on to compute the twist of the DNA, a unitless scalar representing the number of times the strand wound about the helical axis.

Here we are concerned with differences in the properties of these two twists, the step-parameter twist, and the twist of the preceding paragraph, which, because of its connection with the global shape of the helical axis of a closed DNA molecule, a plasmid, for example, we shall refer to as the twist of supercoiling. In the next section we begin by reviewing the definition of the twist of supercoiling, and its well-known connection with the writhing number and the linking number.[12,13] Then, after characterizing two space curves consistent with today’s picture of DNA as a succession of discrete rectangular planes, we go on to describe a method for the computation of the twist of supercoiling for a single base-pair step. We also point out how easy it is to compute the writhe for the case of a closed molecule with an axial curve envisioned as a succession of line segments connecting the origins.

Of particular interest is the difference in properties between the step-parameter twist and the twist of supercoiling. We note that in a relaxed, undeformed configuration of a DNA molecule, the two twists are expected to be close in value for all base-pair steps. However, we find that although translations of the base pairs leave the step-parameter twist unchanged, that is not generally the case for the twist of supercoiling. It, instead, is sensitive to translational distortions that are chiral in nature. To illustrate the point, we compare the values of the two types of twist for certain base-pair steps in DNA wrapped around the core of eight histone proteins in a nucleosome.[14] We find for the chosen base-pair steps an instance in which the local protein-induced deformation of the DNA structure preserves the value of the step-parameter twist but alters the twist of supercoiling.

## THE TWIST OF SUPERCOILING

As we pointed out above, in much of the early theoretical work describing the equilibrium configurations of DNA, the atomistic details of the molecule were ignored, and instead, the structure of the molecule was, in fact, described in terms of two space curves. The tertiary structure of the molecule was represented, as shown in Fig. 1, by the shape of a smooth space curve *C* mirroring the shape of the axis of the double helix.

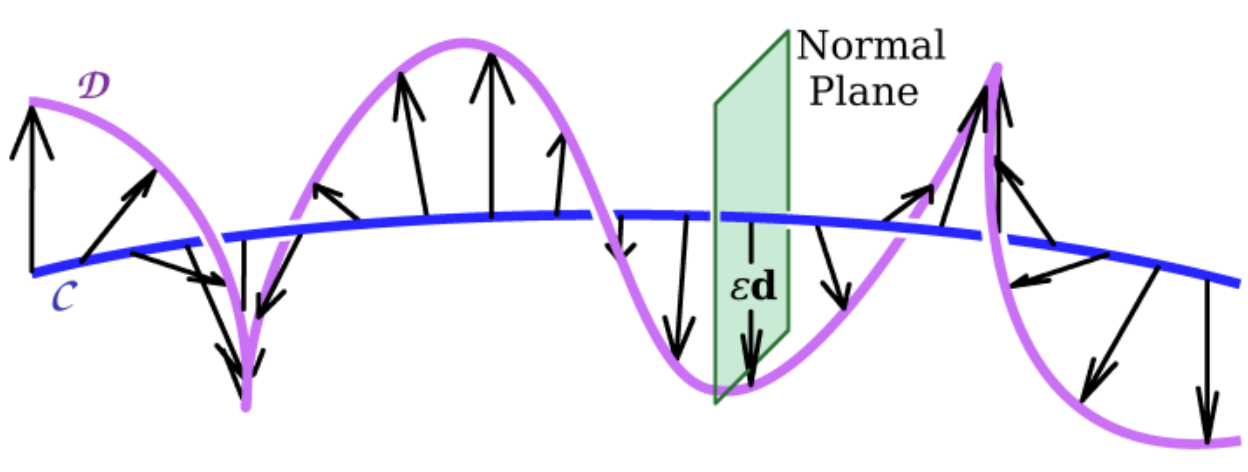

Figure 1: Schematic representation of DNA. The double helical axis is given by curve *C* and one of the helical strands by curve *D*. For purposes of calculation of the twist of *D* about *C*, *D* is to be thought of as being traced out by the head of a vector $\varepsilon\mathbf{d}(s_C)$ everywhere normal to the tangent vector $\mathbf{t}_C(s_C)$.

A second curve *D*, identified with one of the helical strands, was that traced out by the head of a vector $\varepsilon\mathbf{d}(s_C)$, where $\varepsilon$ is a constant, and $\mathbf{d}(s_C)$ is a unit vector normal to the tangent $\mathbf{t}_C(s_C)$ to *C* at a position having an arc length $s_C$ along *C*.

The $\mathbf{t}_C(s_C)$, $\mathbf{d}(s_C)$ pairs associated with two nearby points along *C* having arc lengths $s_C$ and $s_C + ds_C$ are, in general, somewhat rotated with respect to each other. The relative rotational orientation of the two pairs before and after a small change in arc length $ds_C$ is such that there exists a single rotation of the initial pair (at $s_C$) by a small angle about some axis that leads to $\mathbf{t}_C(s_C + ds_C)$ and $\mathbf{d}(s_C + ds_C)$. A vector $d\mathbf{\Omega}$ is defined pointing in the direction of this axis of rotation and having a magnitude equal to the small angle of rotation. The change $d\mathbf{t}_C(s_C)$ and $d\mathbf{d}(s_C)$ that the vectors $\mathbf{t}_C(s_C)$ and $\mathbf{d}(s_C)$ undergo during this change in arc length $ds_C$ is simply the cross product of $d\mathbf{\Omega}$ with the vector itself. That is, if $\mathbf{v}_C(s_C)$ stands for either $\mathbf{t}_C(s_C)$ or for $\mathbf{d}(s_C)$,

$$d\mathbf{v}_C(s_C) = d\mathbf{\Omega} \times \mathbf{v}_C(s_C)\,. \qquad (1)$$

For the case of the tangent, Eq. (1) leads to the equation

$$d\mathbf{\Omega} = \mathbf{t}_C(s_C) \times d\mathbf{t}_C(s_C) + (d\mathbf{\Omega} \cdot \mathbf{t}_C(s_C))\mathbf{t}_C(s_C)\,. \qquad (2)$$

One of the Frenet-Serret equations, the three equations relating the tangent to the principal normal $\mathbf{n}_C(s_C)$ and the binormal $\mathbf{b}_C(s_C)(= \mathbf{t}_C(s_C) \times \mathbf{n}_C(s_C))$, allows us to write $d\mathbf{t}_C(s_C) = \mathbf{n}_C(s_C)\kappa_C(s_C)ds_C$ where $\kappa_C(s_C)$ is the curvature of C. The first term on the right-hand-side of Eq. (2) then becomes

$$\mathbf{t}_C(s_C) \times d\mathbf{t}_C(s_C) = \mathbf{b}_C(s_C)\kappa(s_C)ds_C\,. \qquad (3)$$

The second term in that equation, the one containing the component of $d\mathbf{\Omega}$ along $\mathbf{t}_C(s_C)$, is proportional to the twist density. Replacing $\mathbf{t}_C(s_C)$ by $\mathbf{d}(s_C)$ in Eq. (2) and then taking the projection of the resulting expression for $d\mathbf{\Omega}$ along the tangent shows that

$$d\mathbf{\Omega} \cdot \mathbf{t}_C(s_C) = (\mathbf{d}(s_C) \times d\mathbf{d}(s_C)) \cdot \mathbf{t}_C(s_C)\,. \qquad (4)$$

The twist $T(D,C)$, in units of number of turns, of *D* about a length *l* of *C* is

$$T(D,C) = \left(\frac{1}{2\pi}\right)\int_{s_{C1}}^{s_{C2}} d\mathbf{\Omega}\cdot\mathbf{t}_C(s_C)\,, \tag{5}$$

where $s_{C_2} - s_{C_1} = l$.

If the curves *C*, given by $\mathbf{r}_C(s_C)$, and *D*, given by $\mathbf{r}_D(s_D)$, are closed, the conventional twist is simply related to two other integrals,[12,13] so-called Gauss integrals, the linking number $L(C,D)$

$$L(D,C)=\left(\frac{1}{4\pi}\right)\oint\oint \frac{\mathbf{t}_D(s_D)\times\mathbf{t}_C(s_C)\cdot\left(\mathbf{r}_D(s_D)-\mathbf{r}_C(s_C)\right)}{|\,\mathbf{r}_D(s_D)-\mathbf{r}_C(s_C)|^3}ds_D ds_C \tag{6}$$

and the writhing number $W(C)$, or writhe for short,

$$W(C)=\left(\frac{1}{4\pi}\right)\oint\oint \frac{\mathbf{t}_C(s_C)\times\mathbf{t}_C(s'_C)\cdot\left(\mathbf{r}_C(s_C)-\mathbf{r}_C(s'_C)\right)}{|\left(\mathbf{r}_C(s_C)-\mathbf{r}_C(s'_C)\right)|^3}ds_C ds'_C\,. \tag{7}$$

The linking number is an integer, a topological invariant, equal to the number of times that the curve *D* passes through a surface bounded by *C*.[15] (In computing this sum each pass-through is assigned a value of either +1 or –1 according to a convention consistent with the form of the Gauss integral.) This integer remains unchanged for all distortions in shape of the curves *C* and *D* as long as the curves do not intersect each other during the distortions. The writhe, a property of closed curve *C* alone, is a measure of the chiral distortion of the curve from planarity. Fuller pointed out that its value is also what one would get by averaging, over all orientations of a plane *P*, the sum of the signed self-crossings occurring in the planar curves resulting from the perpendicular projection of *C* on *P*.[10]

The connection between the twist, the writhe, and the linking number mentioned above is given by the well-known equation[9] (For another derivation, see Appendix A.):

$$L(D,C) = W(C) + T(D,C). \tag{8}$$

## THE TWIST OF SUPERCOILING FOR THE MULTISTEP DNA MOLECULE

As we have mentioned, details of the structure of DNA are now more realistically represented, not in terms of smooth space curves but as a sequence of base pairs, each pair of adjacent ones enclosing a base-pair step.[1] The base pair is pictured in Fig. 2 as a rectangular planar slab containing an origin **o** from which a triad of mutually orthogonal unit vectors emanate, a short axis $\mathbf{s}$ , a long axis $\mathbf{l}$, and a normal $\mathbf{n}\,(=\mathbf{s}\times\mathbf{l})$.

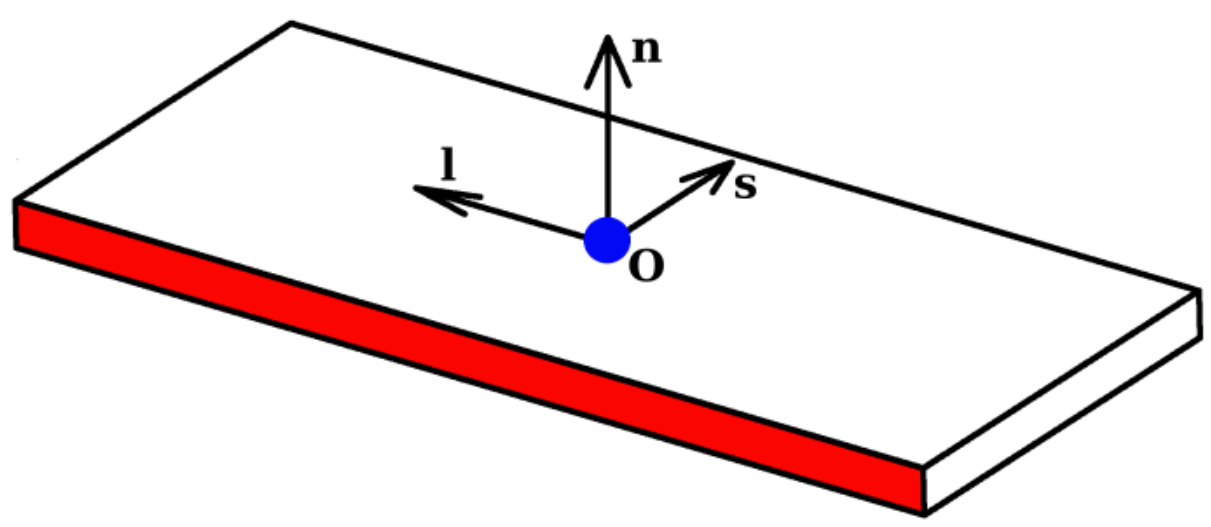

Figure 2: The vectors associated with a base-pair plane: an origin **o**, and a mutually orthogonal triad of unit vectors, the short axis **s**, the long axis **l**, and the normal **n**.

We begin the application of the concept of the twist of supercoiling, described above, to this structure by imagining that there are simply line segments connecting the origins of adjacent base pairs. How then can a smooth curve C be chosen that gives such a picture of the DNA, a picture that seems to show a curve with a discontinuous change in its tangent at each base pair? That is, for each base pair, the $i^{\text{th}}$ for example, the incoming line segment has a unit tangent we call $\mathbf{t}_{(i-1)}$. The tangent of the outgoing segment is $\mathbf{t}_{(i)}$. Both of these vectors are defined in terms of the origins $\mathbf{o}_{i-1}$, $\mathbf{o}_i$, and $\mathbf{o}_{i+1}$ of the $i^{\text{th}}$ base pair and the two base pairs adjoining it

$$\mathbf{t}_{(i-1)} = \frac{\mathbf{o}_i - \mathbf{o}_{i-1}}{|\,\mathbf{o}_i - \mathbf{o}_{i-1}\,|}$$

$$\mathbf{t}_{(i)} = \frac{\mathbf{o}_{i+1} - \mathbf{o}_i}{|\,\mathbf{o}_{i+1} - \mathbf{o}_i\,|}. \tag{9}$$

(Note: Subscripts enclosed in parentheses label base-pair steps and those not enclosed in parentheses label individual base pairs.)

We can envision a limiting process, shown in Fig. 3, in which we start with a curve along which the tangent changes smoothly from $\mathbf{t}_{(i-1)}$ to $\mathbf{t}_{(i)}$ in the vicinity of base pair *i* as one moves along a circular arc lying in a plane spanned by these two vectors, i.e., lying in the plane having as its normal $\mathbf{b}_i$ given by

$$\mathbf{b}_i = \frac{\mathbf{t}_{(i-1)} \times \mathbf{t}_{(i)}}{|\,\mathbf{t}_{(i-1)} \times \mathbf{t}_{(i)}\,|}. \tag{10}$$

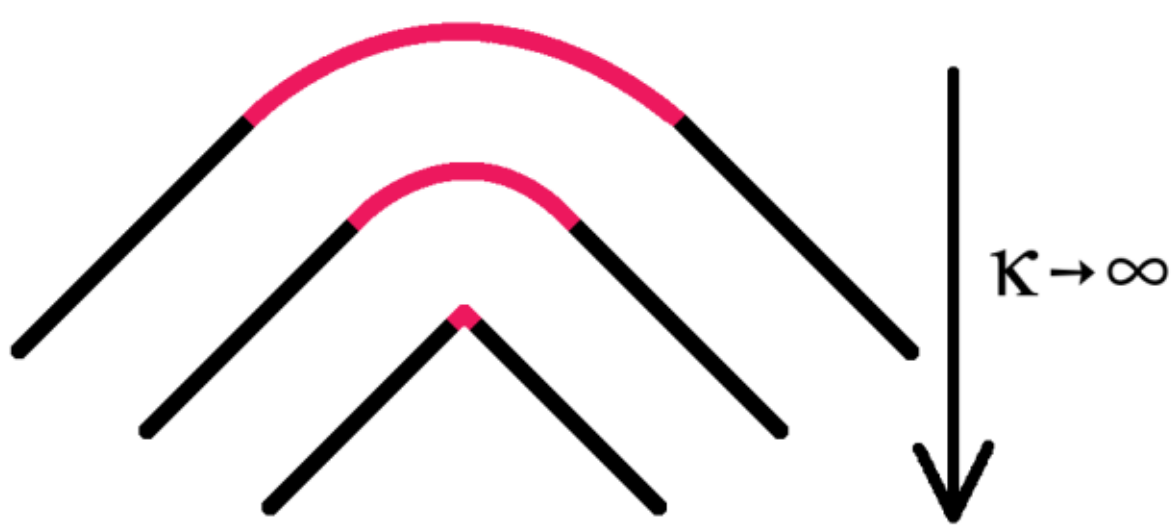

Figure 3: The passage from a smooth curve with circular segments of curvature $\kappa$ to the entirely linear segments connecting the origins of the DNA base pairs.

Then the arc of the circle is allowed to decrease to zero as its curvature increases to infinity, with the tangent vectors $\mathbf{t}_{(i-1)}$ and $\mathbf{t}_{(i)}$ remaining unchanged. This process is repeated at each base pair. Thus, base pair *i*+1 has an incoming tangent $\mathbf{t}_{(i)}$ and an outgoing tangent $\mathbf{t}_{(i+1)}$ determined as in Eqs. (9) from the origins of base pairs *i*+1 and *i*+2. When the tangent becomes $\mathbf{t}_{(i)}$ the curve abruptly becomes a line segment of length $|\,\mathbf{o}_{i+1} - \mathbf{o}_i\,|$ directed along $\mathbf{t}_{(i)}$. At base pair *i*+1 the curvature abruptly becomes infinitely large again but with the tangent still changing smoothly from $\mathbf{t}_{(i)}$ to $\mathbf{t}_{(i+1)}$, along a circular arc lying in a plane with

$$\mathbf{b}_{i+1} = \frac{\mathbf{t}_{(i)} \times \mathbf{t}_{(i+1)}}{|\,\mathbf{t}_{(i)} \times \mathbf{t}_{(i+1)}\,|}. \tag{11}$$

as its normal.

We now can define in greater detail the nature of a single step, the $i^{\text{th}}$. At base pair $i$ curve C for the step begins at that point in the circular arc where the tangent, call it $\tilde{\mathbf{t}}_i$, is midway between $\mathbf{t}_{(i-1)}$ and $\mathbf{t}_{(i)}$ i.e.,

$$\tilde{\mathbf{t}}_i = \frac{\mathbf{t}_{(i-1)} + \mathbf{t}_{(i)}}{|\,\mathbf{t}_{(i-1)} + \mathbf{t}_{(i)}\,|} \, . \tag{12}$$

and ends at the corresponding point at base pair $i$+1, i.e., where the tangent is

$$\tilde{\mathbf{t}}_{i+1} = \frac{\mathbf{t}_{(i)} + \mathbf{t}_{(i+1)}}{|\,\mathbf{t}_{(i)} + \mathbf{t}_{(i+1)}\,|} \, . \tag{13}$$

The unit vector $\mathbf{d}_i^{\mathbf{l}}$ at the beginning of the step we first take to be in the direction of the projection of the long axis $\mathbf{l}_i$ on the plane containing $\mathbf{o}_i$ perpendicular to $\tilde{\mathbf{t}}_i$. The unit vector $\mathbf{d}_{i+1}^{\mathbf{l}}$ at the end of the step points along the projection of $\mathbf{l}_{i+1}$ on the plane containing $\mathbf{o}_{i+1}$ perpendicular to $\tilde{\mathbf{t}}_{i+1}$:

$$\mathbf{d}_i^{\mathbf{l}} = \frac{(\mathbf{l}_i \cdot \mathbf{b}_i)\mathbf{b}_i + (\mathbf{l}_i \cdot \tilde{\mathbf{t}}_i \times \mathbf{b}_i)(\tilde{\mathbf{t}}_i \times \mathbf{b}_i)}{\left|(\mathbf{l}_i \cdot \mathbf{b}_i)\mathbf{b}_i + (\mathbf{l}_i \cdot \tilde{\mathbf{t}}_i \times \mathbf{b}_i)(\tilde{\mathbf{t}}_i \times \mathbf{b}_i)\right|}$$

and (14)

$$\mathbf{d}_{i+1}^{\mathbf{l}} = \frac{(\mathbf{l}_{i+1} \cdot \mathbf{b}_{i+1})\mathbf{b}_{i+1} + (\mathbf{l}_{i+1} \cdot \tilde{\mathbf{t}}_{i+1} \times \mathbf{b}_{i+1})(\tilde{\mathbf{t}}_{i+1} \times \mathbf{b}_{i+1})}{\left|(\mathbf{l}_{i+1} \cdot \mathbf{b}_{i+1})\mathbf{b}_{i+1} + (\mathbf{l}_{i+1} \cdot \tilde{\mathbf{t}}_{i+1} \times \mathbf{b}_{i+1})(\tilde{\mathbf{t}}_{i+1} \times \mathbf{b}_{i+1})\right|}$$

Let us call $\alpha_i^{\mathbf{l}}$ the angle (in radians) that $\mathbf{d}_i^{\mathbf{l}}$ makes with $\mathbf{b}_i$, and $\alpha_{i+1}^{\mathbf{l}}$ the angle that $\mathbf{d}_{i+1}^{\mathbf{l}}$ makes with $\mathbf{b}_{i+1}$, or, more precisely,

$$\cos\alpha_i^{\mathbf{l}} = \mathbf{b}_i \cdot \mathbf{d}_i^{\mathbf{l}}, \qquad \sin\alpha_i^{\mathbf{l}} = \tilde{\mathbf{t}}_i \cdot \mathbf{b}_i \times \mathbf{d}_i^{\mathbf{l}}$$

and (15)

$$\cos\alpha_{i+1}^{\mathbf{l}} = \mathbf{b}_{i+1} \cdot \mathbf{d}_{i+1}^{\mathbf{l}}, \qquad \sin\alpha_{i+1}^{\mathbf{l}} = \tilde{\mathbf{t}}_{i+1} \cdot \mathbf{b}_{i+1} \times \mathbf{d}_{i+1}^{\mathbf{l}}$$

These vectors and angles are pictured in Fig. 4.

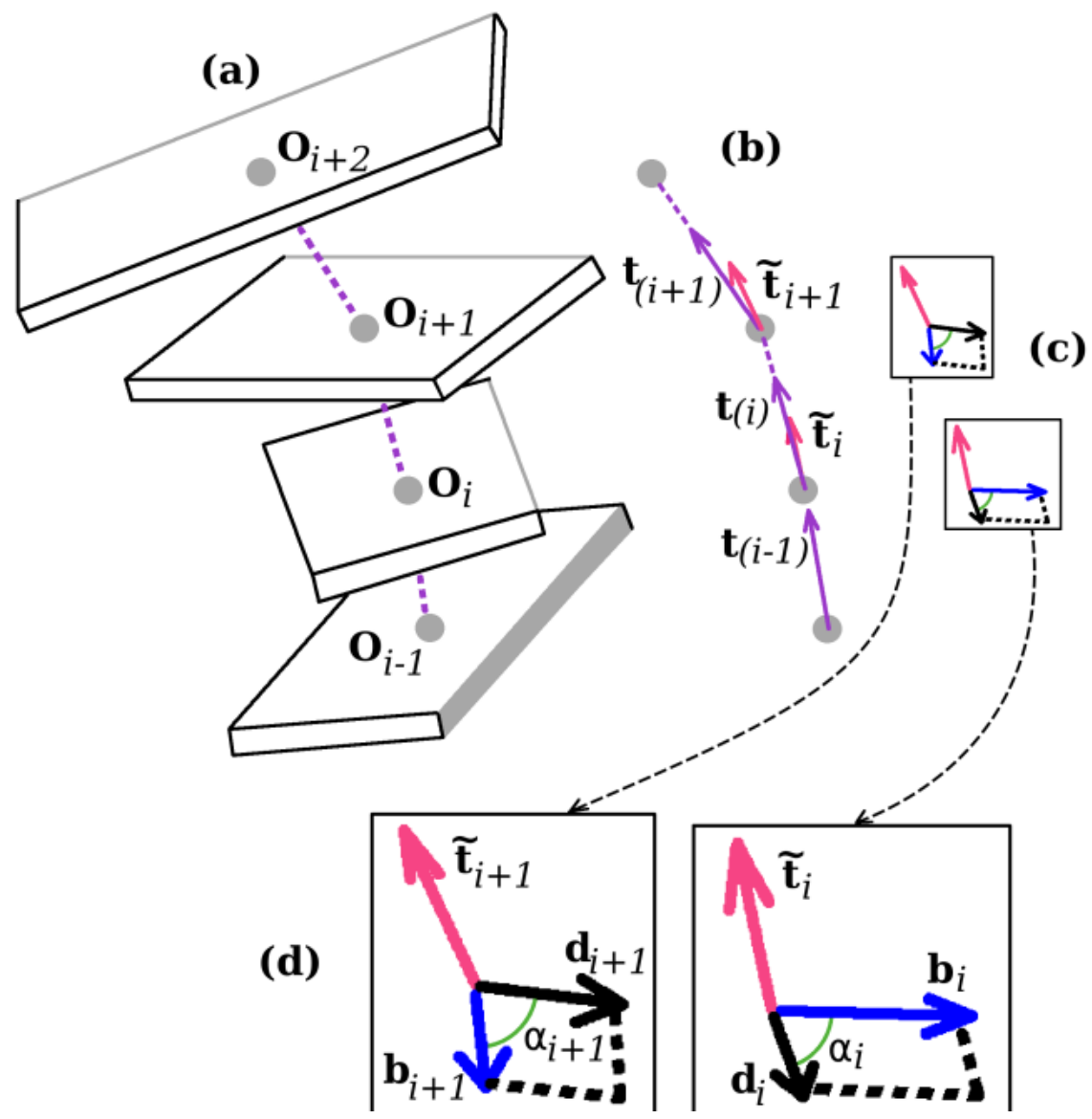

Figure 4: The vectors involved in the calculation of the twist of supercoiling of the DNA base-pair step bounded by the $i^{\text{th}}$ and the $(i+1)^{\text{st}}$ planes. Shown in (a), the four base-pair plane origins needed for the determination of the three vectors depicted in (b), $\mathbf{t}_{(\mathbf{i-1}),}\ \mathbf{t}_{(\mathbf{i}),}$ and $\mathbf{t}_{(\mathbf{i+1})}$, which, in turn, are needed for specifying $\tilde{\mathbf{t}}_i$ and $\tilde{\mathbf{t}}_{i+1}$. These last two are normal to the two planes seen in (c). Each of these planes contains a **d** vector and a **b** vector. Two of the angles needed for the twist calculation as given by Eq. (16) are denoted in (d), an enlargement of (c).

To determine the twist of supercoiling of the $i^{\text{th}}$ base-pair step we carry out five rotations during which $\mathbf{d}_i^{\mathbf{I}}$ is converted to $\mathbf{d}_{i+1}^{\mathbf{I}}$. All of the vectors involved during this process can be thought of as emanating from a single point. For each of the rotations of $\mathbf{d}^{\mathbf{I}}$ the rotational axis, $d\mathbf{\Omega}$, also passing through this point, will either (*a*) lie along a tangent which is not changing its direction, or (*b*) the axis will coincide with one of the two constant **b** vectors. For *a*–type rotations, according to Eq. (5), the twist of supercoiling is simply the angle of rotation while for *b*–type rotations with the axis perpendicular to the tangent, the twist is zero. One: We start with the plane containing $\mathbf{d}_i^{\mathbf{I}}$ and $\mathbf{b}_i$. The normal to this plane is $\tilde{\mathbf{t}}_i$. Now rotate $\mathbf{d}_i^{\mathbf{I}}$ about $\tilde{\mathbf{t}}_i$ until the angle between $\mathbf{d}^{\mathbf{I}}$ and $\mathbf{b}_i$ changes from $\alpha_i^{\mathbf{I}}$ to $\bar{\alpha}_{(i)}^{\mathbf{I}}$ where $\bar{\alpha}_{(i)}^{\mathbf{I}} = (\alpha_i^{\mathbf{I}} + \alpha_{i+1}^{\mathbf{I}})/2$. Two: Rotate $\mathbf{d}^{\mathbf{I}}$ about $\mathbf{b}_i$ until the normal to the plane containing these vectors changes from $\tilde{\mathbf{t}}_i$ to $\mathbf{t}_{(i)}$. At this point the plane contains $\mathbf{b}_{i+1}$ as well as $\mathbf{b}_i$. Three: Rotate $\mathbf{d}^{\mathbf{I}}$ about $\mathbf{t}_{(i)}$ until $\mathbf{d}^{\mathbf{I}}$ makes an angle of $\bar{\alpha}^{\mathbf{l}}_{(i)}$ with $\mathbf{b}_{i+1}$. Four: Rotate $\mathbf{d}^{\mathbf{l}}$ about $\mathbf{b}_{i+1}$ until the normal changes from $\mathbf{t}_{(i)}$ to $\tilde{\mathbf{t}}_{i+1}$. Five: Rotate $\mathbf{d}^{\mathbf{l}}$ about $\tilde{\mathbf{t}}_{i+1}$ until the angle between $\mathbf{d}^{\mathbf{l}}$ and $\mathbf{b}_{i+1}$ changes from $\bar{\alpha}^{\mathbf{l}}_{(i)}$ to $\alpha^{\mathbf{l}}_{i+1}$. The $\mathbf{d}^{\mathbf{l}}$ vector has now become $\mathbf{d}^{\mathbf{l}}_{i+1}$. There is a nonzero twist associated with the *a*–type rotations one, three, and five, since, for each of these steps, the vector $d\mathbf{\Omega}$ is directed along the tangent. Rotations two and four, on the other hand, since the axis of rotation is perpendicular to the tangent, are *b*–type rotations with zero twist. In the case of rotation one and five the twist angle is $\Delta\alpha^{\mathbf{l}}_{(i)}/2$ where $\Delta\alpha^{\mathbf{l}}_{(i)} = \alpha^{\mathbf{l}}_{i+1} - \alpha^{\mathbf{l}}_{i}$. For rotation three the angle of rotation, which we call $\beta_{(i)}$, has as its cosine and sine: $\cos\beta_{(i)} = \mathbf{b}_i \cdot \mathbf{b}_{i+1}$ and $\sin\beta_{(i)} = \mathbf{t}_{(i)} \cdot \mathbf{b}_i \times \mathbf{b}_{i+1}$. The twist of supercoiling of the base-pair step associated with these rotations $T^{\mathbf{l}}_{(i)}$ (in units of number of turns) is thus

$$T^{\mathbf{l}}_{(i)} = \frac{\Delta\alpha^{\mathbf{l}}_{(i)} + \beta_{(i)}}{2\pi}. \tag{16}$$

The beginning and ending **d** vectors in the rotations leading to the twist given by Eq. (16) are those defined in Eqs. (14) and (15), the unit vectors in the direction of the projections of the long axes $\mathbf{l}_i$ and $\mathbf{l}_{i+1}$ onto the planes perpendicular to $\tilde{\mathbf{t}}_i$ and $\tilde{\mathbf{t}}_{i+1}$, respectively.

For an unnicked closed DNA molecule with $n_B$ base pairs, Eq. (8) tells us that

$$\sum_{i=1}^{n_B} T^{\mathbf{l}}_{(i)} = L(D,C) - W(C). \tag{17}$$

If one now were to calculate the $T_{(i)}$'s for the same five rotations but using as the **d**-vectors those derived from the short axes $\mathbf{s}_i$ and $\mathbf{s}_{i+1}$, we would find that for a single step the two twists, call them, $T^{\mathbf{l}}_{(i)}$ and $T^{\mathbf{s}}_{(i)}$, would be somewhat different. The sum $\sum_{i=1}^{n_B} T^{\mathbf{s}}_{(i)}$, however, would have exactly the same value as that obtained before, that given by the right-hand-side of Eq. (17). We define the twist of supercoiling of the step, $T_{(i)}$, as the average of $T^{\mathbf{l}}_{(i)}$ and $T^{\mathbf{s}}_{(i)}$. Clearly, Eq. (17) is also satisfied for $T_{(i)}$'s so defined. The average of $\alpha^{\mathbf{l}}_{i}$ and $\alpha^{\mathbf{s}}_{i}$ will be denoted as $\alpha_i$.

In the calculation of the twist of supercoiling for the case of a DNA molecule with an open helical axis, for the initial step we take $\tilde{\mathbf{t}}_1$ to be in the direction of $\mathbf{t}_{(1)}$, and for the final step we take $\tilde{\mathbf{t}}_{n_B}$ in the direction of $\mathbf{t}_{(n_B-1)}$.

The step-parameter twist of the base pair step can be derived in a similar way. In this case the **d** vectors are the long and short axes themselves. Carrying out the rotations starting with $\mathbf{l}_i$ and ending with $\mathbf{l}_{i+1}$ gives the same value for the twist as starting with $\mathbf{s}_i$ and ending with $\mathbf{s}_{i+1}$. There is thus no need to do the averaging indicated in the definition of the twist of supercoiling. Figure 5 shows the vectors and angles that play a role in the determination of the step-parameter twist.

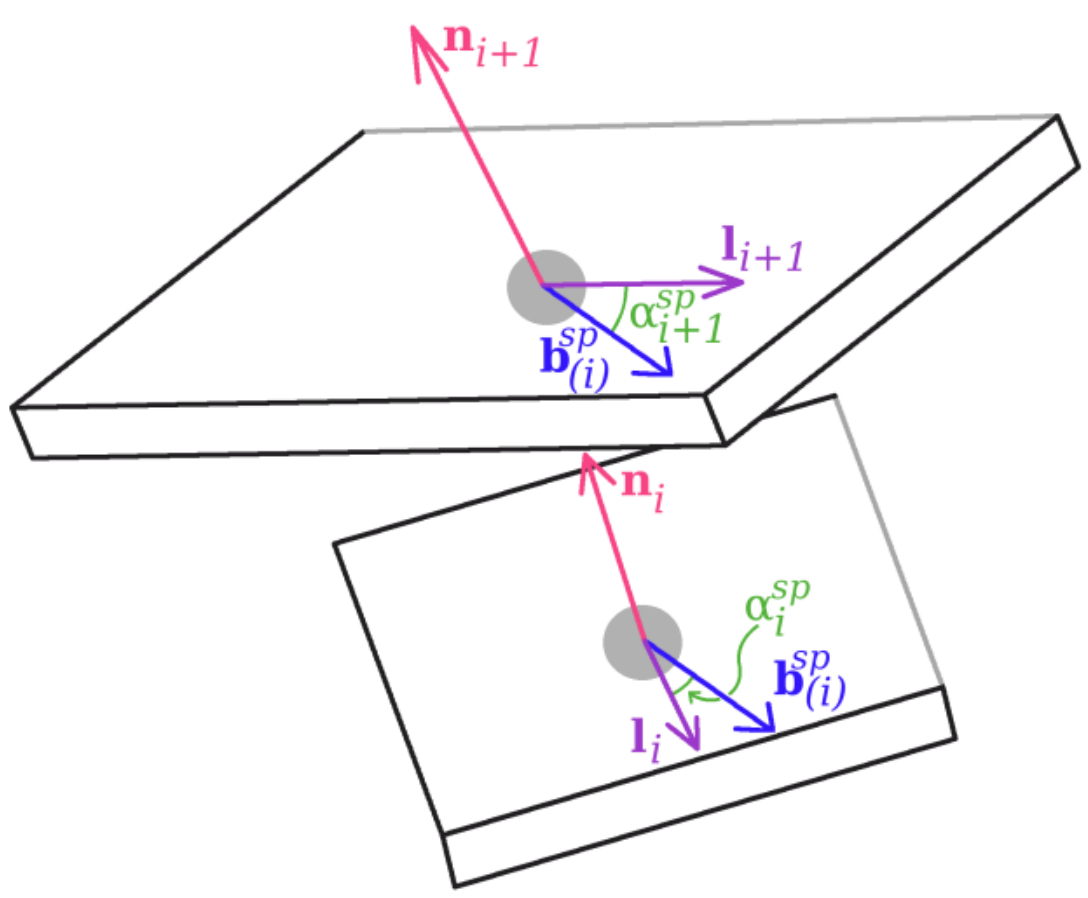

Figure 5: The vectors needed for the calculation of the step-parameter twist of the same step shown in the previous figure. Here knowledge of the direction of the two normals allows the determination of the single vector $\mathbf{b}^{sp}_{(i)}$, which lies in each of the two base-pair planes. Also indicated are the two angles needed for the use of Eq. (18) for the twist calculation.

The analog of $\tilde{\mathbf{t}}_i$ and $\tilde{\mathbf{t}}_{i+1}$ are the normals $\mathbf{n}_i$ and $\mathbf{n}_{i+1}$, and $\mathbf{b}_i$ and $\mathbf{b}_{i+1}$ become the single vector, $\mathbf{b}^{sp}_{(i)} = \mathbf{n}_i \times \mathbf{n}_{i+1} / |\,\mathbf{n}_i \times \mathbf{n}_{i+1}\,|$. The step-parameter twist $T^{sp}_{(i)}$ of the step (in units of number of turns) thus, is, simply,

$$T^{sp}_{(i)} = \frac{\Delta\alpha^{sp}_{(i)}}{2\pi}, \qquad (18)$$

where $\Delta\alpha^{sp}_{(i)}$ is the difference between the angle $\alpha^{sp}_{i+1}$ that $\mathbf{l}_{i+1}$ (or $\mathbf{s}_{i+1}$) makes with $\mathbf{b}^{sp}_{(i)}$ and the angle $\alpha^{sp}_i$ that $\mathbf{l}_i$ (or $\mathbf{s}_i$) makes with $\mathbf{b}^{sp}_{(i)}$.

**THE WRITHE OF THE CLOSED MULTISTEP DNA MOLECULE**

The shape of the closed axial curve of the DNA molecules we are dealing with, consistent with the smooth space curves we have described, is a succession of line segments connecting the origins of the base pairs. A method for computing the writhe of such a segmented closed curve was elegantly put forth many years ago by Levitt.[16] Here we get Levitt's result using a different approach. We first note that for this type of curve, the Gauss integral (Eq. (8)) for the writhe for a molecule with $n_B$ base pairs, and therefore, $n_B$ base pair steps, takes the form of a sum of contributions $w_{(i)(j)}$ of all pairs of base-pair steps,

$$W(C)=\sum_{i,j} w_{(i,j)} \,. \tag{19}$$

with

$$w_{(i,j)} = \left(\frac{1}{2\pi}\right)\iint \frac{\mathbf{t}_{(i)} \times \mathbf{t}_{(j)} \cdot \mathbf{r}_{(i,j)}(s_{(i)},s_{(j)})}{|\,\mathbf{r}_{(i.j)}(s_{(i)},s_{(j)})\,|^3}\, ds_{(i)}ds_{(j)} \tag{20}$$

where

$$\begin{aligned}\mathbf{r}_{(i,j)}(s_{(i)},s_{(j)}) &= \mathbf{r}_{(i)}(s_{(i)}) - \mathbf{r}_{(j)}(s_{(j)}) \\ &= \mathbf{X}_{(i,j)} + \mathbf{t}_{(i)}s_{(i)} - \mathbf{t}_{(j)}s_{(j)}\end{aligned} \tag{21}$$

is a vector which points from a point on the $j^{\text{th}}$ line segment to a point on the $i^{\text{th}}$. The constant vector $\mathbf{X}_{(i,j)}$ is perpendicular to each of the tangents. Its magnitude is thus the distance of closest approach of the segments. Terms involving a base-pair step with itself, and terms involving pairs of adjacent steps are zero.

Given Eq. (21), Eq. (20) can be cast into the form

$$w_{(i,j)} = \left(\frac{1}{2\pi}\right)\int_{s_{(i)}(1)}^{s_{(i)}(2)}\int_{s_{(j)}(1)}^{s_{(j)}(2)} \frac{\mathbf{t}_{(i)} \times \mathbf{t}_{(j)} \cdot \mathbf{X}_{(i,j)}}{|\,\mathbf{X}_{(i,j)} + \mathbf{t}_{(i)}s_{(i)} - \mathbf{t}_{(j)}s_{(j)}\,|^3}\, ds_{(i)}ds_{(j)} \tag{22}$$

where $s_{(i)}(1)$, $s_{(i)}(2)$, $s_{(j)}(1)$, and $s_{(j)}(2)$ denote the arc lengths of the endpoints of the segments.

The three vectors $\mathbf{t}_{(i)}$, $\mathbf{t}_{(j)}$, and $\mathbf{r}_{(i.j)}(s_{(i)}, s_{(j)})$ determine a dihedral angle $\mu_{(i.j)}(s_{(i)}, s_{(j)})$ we define as follows in terms of its sine and cosine

$$\sin\mu_{(i,j)} = -\frac{\mathbf{r}_{(i,j)}\cdot(\mathbf{r}_{(i,j)}\times\mathbf{t}_{(j)})\times(\mathbf{t}_{(i)}\times\mathbf{r}_{(i,j)})}{|\mathbf{r}_{(i,j)}||\mathbf{r}_{(i,j)}\times\mathbf{t}_{(j)}||\mathbf{t}_{(i)}\times\mathbf{r}_{(i,j)}|}$$

$$= -\frac{|\mathbf{r}_{(i,j)}|(\mathbf{r}_{(i,j)}\cdot\mathbf{t}_{(i)}\times\mathbf{t}_{(j)})}{|\mathbf{r}_{(i,j)}\times\mathbf{t}_{(j)}||\mathbf{t}_{(i)}\times\mathbf{r}_{(i,j)}|} \tag{23}$$

$$\cos\mu_{(i,j)} = \frac{(\mathbf{r}_{(i,j)}\times\mathbf{t}_{(j)})\cdot(\mathbf{t}_{(i)}\times\mathbf{r}_{(i,j)})}{|\mathbf{r}_{(i,j)}\times\mathbf{t}_{(j)}||\mathbf{t}_{(i)}\times\mathbf{r}_{(i,j)}|}$$

It follows from its definition that $\mu_{(i.j)}(s_{(i)}, s_{(j)})$ is the angle between the vectors normal to two planes, the one spanned by $\mathbf{r}_{(i.j)}(s_{(i)}, s_{(j)})$ and $\mathbf{t}_{(i)}$, and the one spanned by $\mathbf{r}_{(i.j)}(s_{(i)}, s_{(j)})$ and $\mathbf{t}_{(j)}$.

In Appendix B, it is shown that the integrand in the integral appearing in Eq. (22) is equal to $-\partial^2\mu_{(i,j)}/\partial s_{(i)}\partial s_{(j)}$ so that the contribution to the writhe of base-pair step $i$ and base-pair step $j$, $w_{(i,j)}$, is simply related to the four dihedral angles associated with the endpoints of the steps, i.e.,

$$w_{(i,j)} = \left(\frac{1}{2\pi}\right)(-\mu_{(i,j)}(s_{(i)}(2), s_{(j)}(2)) + \mu_{(i,j)}(s_{(i)}(2), s_{(j)}(1)) - \mu_{(i,j)}(s_{(i)}(1), s_{(j)}(1)) + \mu_{(i,j)}(s_{(i)}(1), s_{(j)}(2))) \tag{24}$$

We note that $w_{(i,j)}$ has the same sign as $\mathbf{t}_{(i)}\times\mathbf{t}_{(j)}\cdot\mathbf{X}_{(i,j)}$.

## COMPARISON OF THE TWIST OF SUPERCOILING AND THE STEP-PARAMETER TWIST

Like $T^{sp}_{(i)}$ of Eq. (18), the twist of supercoiling of a step in the form as expressed in Eq. (16) is independent of the direction of propagation along the DNA molecule. But whereas $T^{sp}_{(i)}$ does not depend at all on the properties of the two steps adjacent to the $i^{\text{th}}$ base-pair steps, $T_{(i)}$ depends on the previous step ($i$-1), and on the following step ($i$+1). In particular, $\tilde{\mathbf{t}}_i$ depends on the origin $\mathbf{o}_{i-1}$ and $\tilde{\mathbf{t}}_{i+1}$ depends on $\mathbf{o}_{i+1}$. For Eq. (17) to be satisfied for unnicked closed DNA molecules, the twist of supercoiling must depend on the way the tangent vectors are changing. In the case of the step-parameter twist, on the other hand, the sum $\sum_{i=1}^{n_B} T^{sp}_{(i)}$ has no particular significance.

Because in relaxed DNA, the normals of the base pairs and the associated $\tilde{\mathbf{t}}$ vectors are approximately collinear, for each step we expect the two twists to be close in value. When the base pairs are forced to undergo translations of a chiral nature, however, unlike the step-parameter twist, which is unaffected by pure translations, we expect the twist of supercoiling to change.

We illustrate this by considering a model structure suggested by an examination of a particular three-step section of nucleosomal DNA. Table I shows the twist of supercoiling and the step-parameters of three adjacent base-pair steps, 37, 38, and 39, derived from the X-ray measurements carried out on this molecule.[14]

| Step | Shift | Slide | Rise | Tilt | Roll | $T^{sp}_{(i)}$ | $T_{(i)}$ |
|---|---|---|---|---|---|---|---|
| 37 | -0.18 | -0.18 | 3.02 | -0.3 | -0.7 | 28.6 | 29.6 |
| 38 | -0.43 | 2.58 | 3.25 | -2.3 | -18.4 | 50.0 | 45.6 |
| 39 | 0.95 | 0.86 | 3.47 | 2.0 | 7.0 | 28.5 | 28.4 |

Table I: Step parameters and the twist of supercoiling of steps 37-38 of the x-ray crystal structure of a nucleosome core particle. The unit of each of the three translational parameters is angstroms, and of the angular parameters is degrees.

Before discussing the implications of Table I, we review the meaning of the five step parameters Shift, Slide, Rise, Tilt, and Roll. (An expression for the Twist, here called $T^{sp}_{(i)}$, was derived in a preceding section, Eq. (18).) One, first of all, associates with each step a virtual base-pair frame characterized by three mutually orthogonal unit vectors, a normal $\bar{\mathbf{n}}$ pointing in the direction of $\mathbf{n}_i + \mathbf{n}_{i+1}$, a long axis $\bar{\mathbf{l}}$ making an angle $\bar{\alpha}^{sp}_{(i)}$, the average of $\alpha^{sp}_{i}$ and $\alpha^{sp}_{i+1}$, with $\mathbf{b}^{sp}_{(i)}$, which also lies in this plane, and a short axis $\bar{\mathbf{s}}$ given by $\bar{\mathbf{l}} \times \bar{\mathbf{n}}$. Then the shearing translations, shift and slide, are given by the projection of $\mathbf{o}_{i+1} - \mathbf{o}_{i}$ on $\bar{\mathbf{s}}$ and on $\bar{\mathbf{l}}$, respectively, and the rise is given by the projection of $\mathbf{o}_{i+1} - \mathbf{o}_{i}$ on $\bar{\mathbf{n}}$. Tilt is the angle between the normals of the two base pairs defining the step multiplied by the sine of $\bar{\alpha}^{sp}_{(i)}$ and the roll is the same angle multiplied by the cosine of $\bar{\alpha}^{sp}_{(i)}$. Each of the six step parameters is associated with a base-pair orientation in Fig. 6.

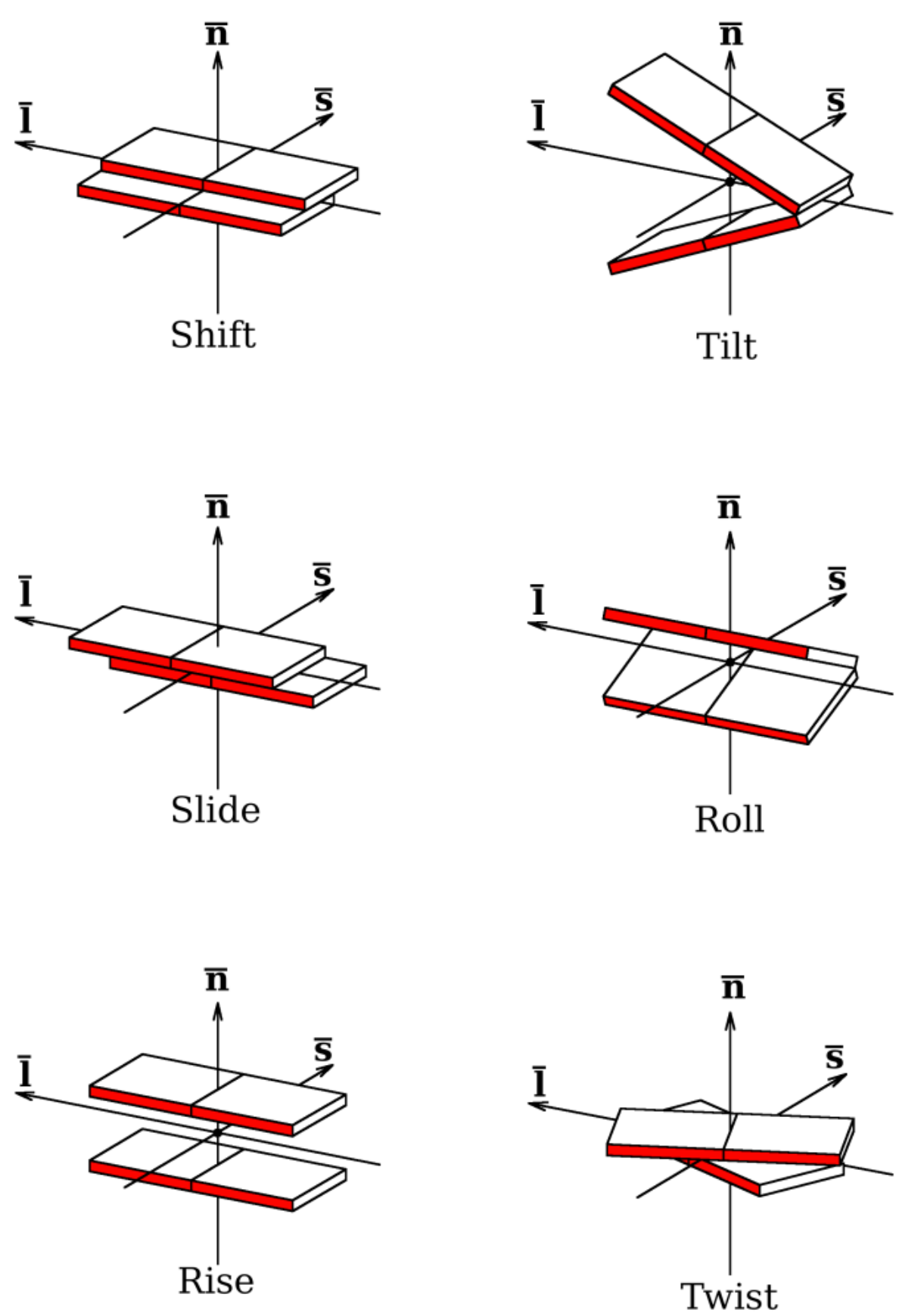

Figure 6: Pictorial definitions of the six rigid-body parameters used to describe the geometry of a base-pair step. The short axis, long axis, and normal of the virtual mid-plane of the step are denoted by $\bar{\mathbf{s}}$, $\bar{\mathbf{l}}$, and $\bar{\mathbf{n}}$. In each of the cases, the image illustrates a positive value of the designated parameter.

Note that in Table I step 38 is characterized by a roll of $-18°$, a slide of 2.6 Å, and a relatively small tilt and shift. This is consistent with a concerted bending and shearing deformation underlying the superhelical folding of nucleosomal DNA.[17] The twist of supercoiling for each of the three steps is also indicated. For step 38 the twist of supercoiling is notably less than the step-parameter twist, unlike the two adjacent steps where the two twists are close in value.

To show how differences in values of the step-parameter twist and the twist of supercoiling can arise, the simple model we consider is depicted in Fig. 7.

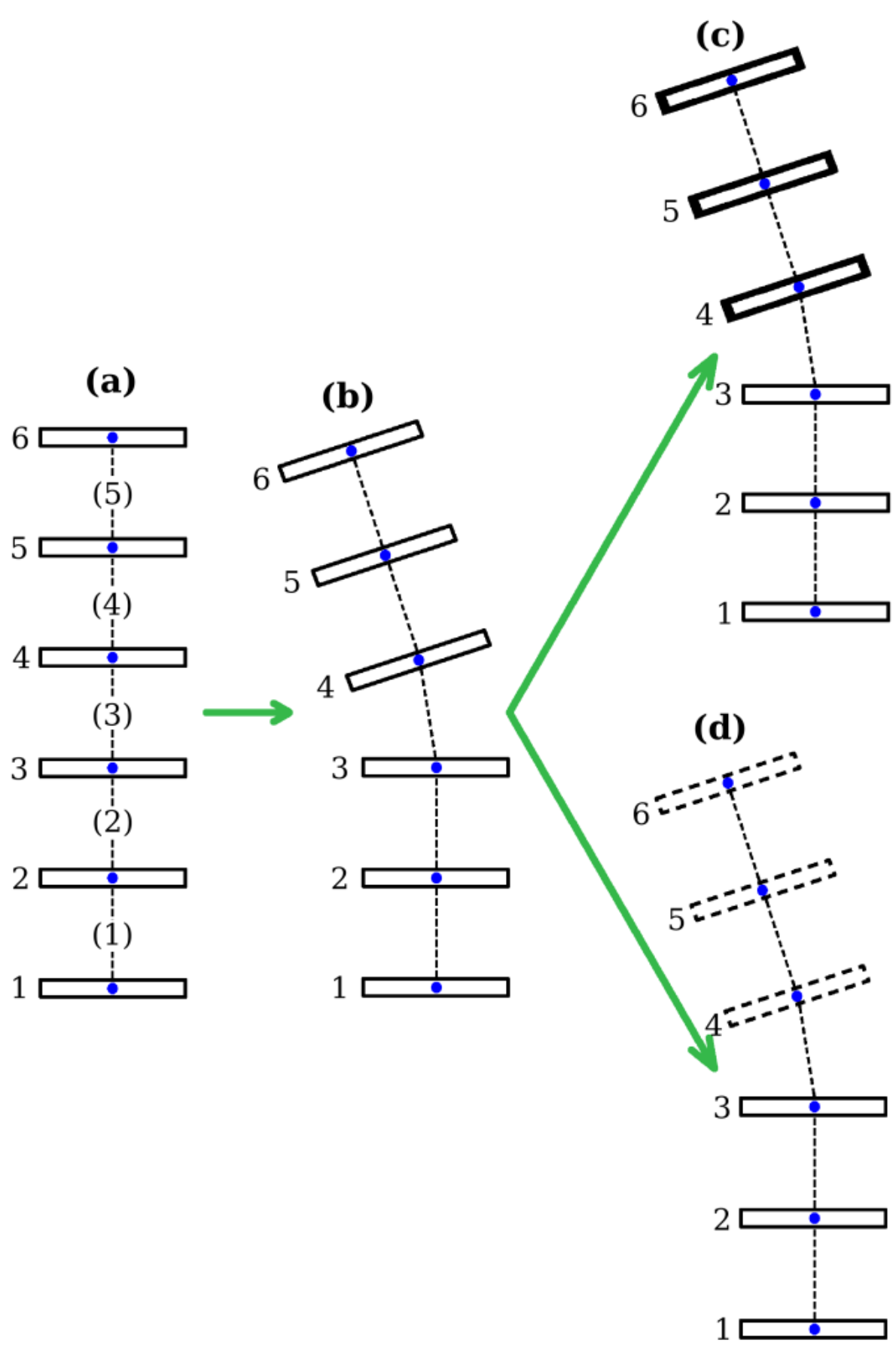

Figure 7: Construction of a model DNA structure characterized by a chiral deformation. In (a) six equally spaced and parallel base-pair planes are shown having their origins lying on a line. (b) depicts the structure after the bend described in the text is introduced. The six origins are still coplanar, and the viewing direction is chosen to be normal to this plane. A translation of base pairs 4, 5, and 6 as a single unit along

the viewing direction, depending on the direction of the motion, results either in a structure with a right-handed jog ((c)) or one with a left-handed jog ((d)).

At the start, in Fig. 7(a), there are six base pairs with their origins collinear and equally spaced. In addition, the base-pair planes are oriented so that their normals coincide with the **t**-vectors. For each of the five steps the rise is simply equal to the distance between the origins, and the step-parameter twist and the twist of supercoiling are equal in value. Also, for each of the steps, aside from the rise and the step-parameter twist, the other step parameters are zero. The line of base pairs 4,5, and 6 is now rotated by some angle about an axis lying in base pair 3 and passing through the origin of that base pair, and then by the same angle about an axis lying in base pair 4 and passing through the origin of that base pair. The effect of the rotation is to change the values of the roll and tilt of step 3. All of the other step parameters are left unchanged. Neither do any of the twists of supercoiling of the steps change. After this bend is introduced, the origins of the base pairs are no longer collinear, but they all do remain coplanar. Figure 7(b) shows the new structure with the viewing direction chosen to be normal to the plane on which the origins lie. Finally, a translation of base pairs 4, 5, and 6 as a unit in each of the two directions perpendicular to the plane on which their origins had been lying gives rise to the two chiral structures, a structure with a right-handed jog shown in Fig. 7(c), and one with a left-handed jog shown in Fig. 7(d). The final shearing motion alters just the slide and the shift of step 3 in the case of the step parameters, and, in fact, the ratio of the roll to the tilt equals the ratio of the slide to the shift. But also accompanying this introduction of structural chirality is a change in the twists of supercoiling of steps 2, 3, and 4. It is this that we now examine in some more detail.

According to Eq. (16), we can write for the twist of supercoiling for each of the three central steps of Fig. 7(c) or 7(d)

$$T_{(i)} = \frac{\alpha_{i+1} - \alpha_i + \beta_{(i)}}{2\pi} \tag{25}$$

with $i$ = 2, 3, 4. The sum of these twists of the steps in question is, therefore,

$$\sum_{i=2}^{4} T_{(i)} = \frac{\alpha_5 - \alpha_2 + \sum_{i=2}^{4} \beta_{(i)}}{2\pi}. \tag{26}$$

Since for the final structures of this model all of the vectors $\mathbf{b}_{(i)}^{sp}$ are unidirectional and the vectors $\mathbf{b}_2$ and $\mathbf{b}_5$ coincide with $\mathbf{b}_2^{sp}$ and $\mathbf{b}_4^{sp}$, it follows that

$$\sum_{i=2}^{4} T_{(i)}^{sp} = \frac{\alpha_5^{sp} - \alpha_2^{sp}}{2\pi} \tag{27}$$

and that $\alpha_2 = \alpha_2^{sp}$ and $\alpha_5 = \alpha_5^{sp}$. Thus

$$\sum_{i=2}^{4} T_{(i)} - \sum_{i=2}^{4} T_{(i)}^{sp} = \frac{\sum_{i=2}^{4} \beta_{(i)}}{2\pi}. \tag{28}$$

That is, for this model, the difference in the sums of twist of supercoiling and the step-parameter twist for the three central steps depends only on the rise, roll, and slide. In particular, the difference does not depend on the values of the step-parameter twists originally assigned in the linear structure of Fig. 7(a). It is also found that the values of the differences $T_{(i)} - T_{(i)}^{sp}$ for each of the three steps are only mildly dependent on the step-parameter twists of the steps. Furthermore, the value of the twist difference when $i$ = 2 or 4 is much smaller than it is for the case of the middle step, $i$ = 3. With a roll, slide, and rise of step 3 of $-18°$, 2.6 Å, and 3.4 Å, respectively, and with zero tilt and shift, the model gives, for the case of the case of the left-handed jog, in units of degrees $\sum_{i=2}^{4} T_{(i)} - \sum_{i=2}^{4} T_{(i)}^{sp} = -6.10°$. Also, when $T_{(2)}^{sp} = 29°$, $T_{(3)}^{sp} = 50°$, and $T_{(4)}^{sp} = 29°$, we find that $T_{(2)} - T_{(2)}^{sp} = -0.01°$, $T_{(3)} - T_{(3)}^{sp} = -6.08°$, and $T_{(4)} - T_{(4)}^{sp} = -0.01°$.

The model structure with the left-handed jog, Fig. 7(d), is seen to have properties qualitatively similar to those shown by the nucleosomal steps 37, 38, and 39. The corresponding twist differences for the nucleosome are $\sum_{i=37}^{39} T_{(i)} - \sum_{i=37}^{39} T_{(i)}^{sp} = -3.7°$, $T_{(37)} - T_{(37)}^{sp} = 0.9°$, $T_{(38)} - T_{(38)}^{sp} = -4.4°$, and $T_{(39)} - T_{(39)}^{sp} = -0.2°$.

**ACKNOWLEDGMENTS**

We thank Luke Czapla for valuable discussions. The U.S. Public Health Service under research grant GM34809 and instrumentation grant RR022375 has generously supported this work.

**Appendix A**

We begin the derivation of Eq. (8) by expressing the connection between curve *D* and curve *C*, by writing for $\mathbf{r}_{\mathrm{D}}$,

$$\mathbf{r}_{\mathrm{D}}(s_{\mathrm{D}}) = \mathbf{r}_{\mathrm{C}}(s'_{\mathrm{C}}) + \varepsilon \mathbf{d}(s'_{\mathrm{C}}), \tag{A-1}$$

where $\varepsilon$ is a constant and $\mathbf{d}(s'_{\mathrm{C}})$ is a unit vector perpendicular to the tangent $\mathbf{t}_{\mathrm{C}}(s'_{\mathrm{C}})$ to curve *C*, i.e., $\mathbf{d}(s'_{\mathrm{C}}) \cdot \mathbf{t}_{\mathrm{C}}(s'_{\mathrm{C}}) = 0.$ Equation (A-1) implies that the unit tangent $\mathbf{t}_{\mathrm{D}}(s_{\mathrm{D}})$ to curve *D* is of the form

$$\mathbf{t}_{\mathrm{D}}(s_{\mathrm{D}}) = k\left(\mathbf{t}_{\mathrm{C}}(s'_{\mathrm{C}}) + \varepsilon \frac{d\mathbf{d}(s'_{\mathrm{C}})}{ds'_{\mathrm{C}}}\right) \tag{A-2}$$

where *k* is the reciprocal of the magnitude of the vector contained in the parentheses. These two expressions are substituted into the integrand appearing in the Gauss integral form for the linking number, Eq. (6). We then allow $\varepsilon$ to approach zero. Beyond a certain point in this limiting process curve *C* and curve *D* no longer can intersect each other, and thereafter the linking number remains unchanged.[10] The factor *k* approaches one as $\varepsilon$ approaches zero. We also find that, in this limit, the terms in the integrand that are linear in $\varepsilon$ are zero. Thus one can write for the linking number

$$L(D,C) = W(C) + \lim_{\varepsilon \to 0}\left(\frac{\varepsilon^2}{4\pi}\right)\oiint \frac{\mathbf{d}(s_{\mathrm{C}}) \times \dfrac{d\mathbf{d}(s_{\mathrm{C}})}{ds_{\mathrm{C}}} \cdot \mathbf{t}_{\mathrm{C}}(s'_{\mathrm{C}})}{\left(|\,\mathbf{r}_{\mathrm{C}}(s_{\mathrm{C}}) - \mathbf{r}_{\mathrm{C}}(s'_{\mathrm{C}})\,|^2 + \varepsilon^2\right)^{3/2}}\, ds'_{\mathrm{C}} ds_{\mathrm{C}} \tag{A-3}$$

where $W(\mathrm{C})$ is the writhing number as given in Eq. (7). Furthermore, because

$$\begin{aligned}\lim_{\varepsilon \to 0}\frac{\varepsilon^2}{\left(|\,\mathbf{r}_{\mathrm{C}}(s_{\mathrm{C}}) - \mathbf{r}_{\mathrm{C}}(s'_{\mathrm{C}})\,|^2 + \varepsilon^2\right)^{3/2}} &= \lim_{\varepsilon \to 0}\frac{\varepsilon^2}{\left(\left(s_{\mathrm{C}} - s'_{\mathrm{C}}\right)^2 + \varepsilon^2\right)^{3/2}} \\ &= 2\,\delta(s_{\mathrm{C}} - s'_{\mathrm{C}}),\end{aligned} \tag{A-4}$$

where $\delta(s_{\mathrm{C}} - s'_{\mathrm{C}})$ is a Dirac delta function, the integration over $s'_{\mathrm{C}}$ in the second term of Eq. (A-3) can be readily carried out:

$$\lim_{\varepsilon\to 0}\left(\frac{\varepsilon^2}{4\pi}\right)\oint\oint \frac{\mathbf{d}(s_\mathrm{C})\times\frac{d\mathbf{d}(s_\mathrm{C})}{ds_\mathrm{C}}\cdot\mathbf{t}_\mathrm{C}(s'_\mathrm{C})}{\left(|\mathbf{r}_\mathrm{C}(s_\mathrm{C})-\mathbf{r}_\mathrm{C}(s'_\mathrm{C})|^2+\varepsilon^2\right)^{3/2}}\,ds'_\mathrm{C}ds_\mathrm{C}$$

$$=\left(\frac{1}{2\pi}\right)\oint\oint \delta(s_\mathrm{C}-s'_\mathrm{C})\,\mathbf{d}(s_\mathrm{C})\times\frac{d\mathbf{d}(s_\mathrm{C})}{ds_\mathrm{C}}\cdot\mathbf{t}_\mathrm{C}(s'_\mathrm{C})\,ds'_\mathrm{C}ds_\mathrm{C} \qquad \text{(A-5)}$$

$$=\left(\frac{1}{2\pi}\right)\oint \mathbf{d}(s_\mathrm{C})\times\frac{d\mathbf{d}(s_\mathrm{C})}{ds_\mathrm{C}}\cdot\mathbf{t}_\mathrm{C}(s_\mathrm{C})\,ds_\mathrm{C}.$$

This completes the proof of Eq. (8).

**Appendix B**

We begin by deriving an expression for $-\partial\mu_{(i,j)}/\partial s_{(j)}$. Since, for example,

$$\frac{\partial\cot\mu}{\partial s_{(j)}}=\frac{d\cot\mu}{d\mu}\frac{\partial\mu}{\partial s_{(j)}}, \qquad \text{(B-1)}$$

we see that

$$-\frac{\partial\mu}{\partial s_{(j)}}=\sin^2\mu\frac{\partial\cot\mu}{\partial s_{(j)}}. \qquad \text{(B-2)}$$

Given the definition of the dihedral angle (Eqs. (23)), and the fact that $\mathbf{t}_{(i)}\times\mathbf{t}_{(j)}\cdot\mathbf{r}_{(i,j)}$ is independent of $s_{(i)}$ and $s_{(j)}$, this last equation can be rewritten as

$$-\frac{\partial\mu}{\partial s_{(j)}}=-\frac{|\mathbf{r}_{(i,j)}|^2\left(\mathbf{t}_{(i)}\times\mathbf{t}_{(j)}\cdot\mathbf{r}_{(i,j)}\right)}{|\mathbf{r}_{(i,j)}\times\mathbf{t}_{(j)}|^2|\mathbf{t}_{(i)}\times\mathbf{r}_{(i,j)}|^2} \times\frac{\partial}{\partial s_{(j)}}\left(\frac{(\mathbf{r}_{(i,j)}\times\mathbf{t}_{(j)})\cdot(\mathbf{t}_{(i)}\times\mathbf{r}_{(i,j)})}{|\mathbf{r}_{(i,j)}|}\right) \qquad \text{(B-3)}$$

After noting that the explicit dependence of $\mathbf{r}_{(i,j)}$ on $s_{(i)}$ and $s_{(j)}$ as given in Eq. (21) leads to the fact that

$$|\mathbf{r}_{(i,j)}|^2 = |\mathbf{X}_{(i,j)}|^\mathbf{2} + s_{(i)}^2 + s_{(j)}^2 - 2(\mathbf{t}_{(i)} \cdot \mathbf{t}_{(j)}) s_{(i)} s_{(j)}$$

$$\mathbf{t}_{(i)} \times \mathbf{t}_{(j)} \cdot \mathbf{r}_{(i,j)} = \mathbf{t}_{(i)} \times \mathbf{t}_{(j)} \cdot \mathbf{X}_{(i,j)}$$

$$|\mathbf{t}_{(i)} \times \mathbf{r}_{(i,j)}|^2 = |\mathbf{X}_{(i,j)}|^\mathbf{2} + |\mathbf{t}_{(i)} \times \mathbf{t}_{(j)}|^2 \; s_{(j)}^2 \tag{B-4}$$

$$|\mathbf{r}_{(i,j)} \times \mathbf{t}_{(j)}|^2 = |\mathbf{X}_{(i,j)}|^\mathbf{2} + |\mathbf{t}_{(i)} \times \mathbf{t}_{(j)}|^2 \; s_{(i)}^2$$

$$(\mathbf{r}_{(i,j)} \times \mathbf{t}_{(j)}) \cdot (\mathbf{t}_{(i)} \times \mathbf{r}_{(i,j)}) = -\left( (\mathbf{t}_{(i)} \cdot \mathbf{t}_{(j)}) \, |\mathbf{X}_{(i,j)}|^\mathbf{2} + |\mathbf{t}_{(i)} \times \mathbf{t}_{(j)}|^2 \; s_{(i)} s_{(j)} \right)$$

We find, after performing the indicated differentiation with respect to $s_{(j)}$ in Eq. (B-3) and simplifying the resulting expression, that

$$-\frac{\partial \mu}{\partial s_{(j)}} = \frac{\left(\mathbf{t}_{(i)} \times \mathbf{t}_{(j)} \cdot \mathbf{X}_{(i,j)}\right) \; (s_{(i)} - (\mathbf{t}_{(i)} \cdot \mathbf{t}_{(j)}) s_{(j)})}{(|\mathbf{X}_{(i,j)}|^\mathbf{2} + |\mathbf{t}_{(i)} \times \mathbf{t}_{(j)}|^2 \; s_{(j)}^2) \, |\mathbf{r}_{(i,j)}|} \; . \tag{B-5}$$

Differentiating Eq. (B-5) again, this time with respect to $s_{(i)}$, yields

$$-\frac{\partial^2 \mu_{(i,j)}}{\partial s_{(i)} \partial s_{(j)}} = \frac{\mathbf{t}_{(i)} \times \mathbf{t}_{(j)} \cdot \mathbf{X}_{(i,j)}}{|\mathbf{r}_{(i.j)}|^3} \; . \tag{B-6}$$